# An Efficient Sufficient Dimension Reduction Method for Identifying Genetic Variants of Clinical Significance

**Momiao Xiong**
Division of Biostatistics
The University of Texas School of Public Health
Houston, TX 77030
momiao.xiong@uth.tmc.edu

**Long Ma**
Division of Biostatistics
The University of Texas School of Public Health
Houston, TX 77030
marlone.zj@gmail.com

**Abstract**
Fast and cheaper next generation sequencing (NGS) technologies will generate unprecedentedly massive (thousands or even ten thousands of individuals) and highly-dimensional (up to hundreds of millions) genomic and epigenomic variation data. In the near future, a routine part of medical record will include the sequenced genomes. A fundamental question is how to efficiently extract genomic and epigenomic variants of clinical utility which will provide information for optimal wellness and interference strategies. Traditional paradigm for identifying variants of clinical validity is to test association of the variants. However, significantly associated genetic variants may or may not be usefulness for diagnosis and prognosis of diseases. Alternative to association studies for finding genetic variants of predictive utility is to systematically search variants that contain sufficient information for phenotype prediction. To achieve this, we introduce concepts of sufficient dimension reduction (SDR) and coordinate hypothesis which project the original high dimensional data to very low dimensional space while preserving all information on response phenotypes. We then formulate clinically significant genetic variant discovery problem into sparse SDR problem and develop algorithms that can select significant genetic variants from up to or even ten millions of predictors with the aid of dividing SDR for whole genome into a number of sub-SDR problems defined for genomic regions. The sparse SDR is in turn formulated as sparse optimal scoring problem, but with penalty which can remove row vectors from the basis matrix. To speed up computation, we develop the modified alternating direction method for multipliers to solve the sparse optimal scoring problem which can easily be implemented in parallel. To illustrate its application, the proposed method is applied to simulation data and the NHLBI's Exome Sequencing Project (ESP) dataset as well as the TCGA dataset.

## Introduction

Purpose of this paper is to formulate clinically significant genetic variant discovery problem into sparse SDR problem and develop algorithms that can select significant genetic variants from up to millions of predictors. To achieve this, we first show that SDR for whole genome can be partitioned into a number of sub-SDR problems defined for divided genomic regions. Then, similar to Wang and Zhu's approach, we formulate the sparse SDR into sparse optimal scoring problem, but with penalty which can remove row vectors from the basis matrix. Since large-scale discovery of genetic variants may involve millions of genetic variants, solving large sparse optimal scoring problem requires heavy computation. To speed up computation, we apply the alternating direction method for multipliers which can easily be implemented in parallel. To illustrate its application, the proposed method is applied to simulation data and the NHLBI's Exome Sequencing Project (ESP) dataset and TCGA dataset.

## Methods

### Sufficient Dimension Reduction

Throughout the paper we consider continuous phenotype (response variable) and regression. In other words, we will focus on quantitative trait analysis. However, all discussed concepts can be extended to binary response variable and classification. Let $Y$ be a univariate response variable (phenotype) and $X$ be a $p$ dimensional vector of predictors (genotypes for genetic variants). Since dimension of genomic variation is extremely high, to reduce the impact of noise and irrelevant predictors, dimension reduction is a powerful tool for quantitative trait analysis and regression. Dimension reduction is to identify the best linear subspace that that best preserves information relevant to a regression (Nilsson et al. 2007). Dimension reduction consists of unsupervised dimension reduction and supervised dimension reduction. Principal component analysis (PCA) is a typical method for unsupervised dimension reduction which projects predictor data onto a linear space without response information. Supervised dimension reduction is to discover the best subspace that maximally reduces the dimension of the input while preserving the information necessary to predict the response variable. The current popular supervised dimension reduction method is SDR which aims to find a linear subspace S such that the response Y is conditionally independent of the covariate vector X, given the projection of X on S:

$$Y \perp X \mid P_S X , \qquad (1)$$

where $\perp$ indicates independence and $P_S$ represents a projection on $S$. In other words, all information of $X$ about $Y$ is contained in the space $S$. The subspace S is referred to as a dimension reduction subspace. The subspace $S$ may not be unique. To uniquely describe dimension reduction subspace, we introduce central subspace (CS) that is defined the intersections of all reduction subspaces $S$ satisfying conditional independence assertion. The CS is denoted by $S_{Y|X}$.

Many methods have been developed for identifying CS. A popular sliced inverse regression for identifying the basis vector in the CS is to solve the following eigenequation:

$$\text{cov}(E(X - E(X)|Y))\beta = \lambda_x \Sigma \beta, \qquad (2)$$

where $\lambda_x$ are eigenvalues, and $\beta$ is an eigenvector, respectively. Solutions to eigenequation (2) yields the basis matrices $B = [\beta_1,...,\beta_k]$ for $S_{Y|X}$.

## Sparse SDR by Alternative Direction Method of Multipliers

The eigenvalue problem can also been formulated as a constrained optimization problem (Chen and Li 1998; Wang and Zhu 2013):

$$\min_{\theta_i \in R^k, b_i \in R^P} ||Y\theta_i - X\beta_i||^2 \qquad (3)$$
$$\text{s.t.} \quad \theta_i^T Y^T Y \theta_i = 1, \theta_i^T Y^T Y \theta_j = 0, j = 1,...i-1, i = 1,...d,$$

where $Y = [\phi(y_1),...,\phi(y_n)]^T$.

To develop sparse SDR that can simultaneously reduce the dimension and the number of predictors, we first introduce a coordinated-independent penalty function. Let $B = [\beta_1,...,\beta_i] = [\beta_1^{*T},...,\beta_p^{*T}], i = 1,...,d$ be a $p \times i$ matrix which forms the basis matrix of the CS. We introduce the following penalty function to penalize the variable in all reduction directions toward zero (Chen et al. 2010): $\rho(B) = \sum_{l=1}^{p} \lambda_l \sqrt{\beta_l^{*T}\beta_l^*} = \sum_{l=1}^{p} \lambda_l |\beta_l^*|_2$. For simplicity of computation, we define $\lambda_l = \lambda ||\beta_l^*||_2^{-r}$, $r > 0$ is a pre-specified parameter. We can chose $r = 0.5$ as Chen (2010) suggested.

After introducing the penalty function, the sparse version of optimal scoring problem (3) for penalizing the variable can be defined as

$$\min_{\theta_i \in R^k, \beta_i \in R^P} ||Y\theta_i - X\beta_i||_2^2 + \lambda \sum_{l=1}^{p} ||\beta_l^*||_2^{1-r}$$
$$\text{s.t.} \quad \theta_i^T Y^T Y \theta_i = 1, \theta_i^T Y^T Y \theta_j = 0, j < i, i = 1,...d,$$

This is a bi-convex problem. It is convex in $\theta$ for each $\beta$ and convex in $\beta$ for each $\theta$. It can be solved by a simple iterative algorithm. The iterative process consists of two steps: (1) for fixed $\theta_i$ we optimize with respect to $\beta_i$ and for fixed $\beta_i$ we optimize with respect to $\theta_i$. The algorithms are given bellow.

Step 1: Initialization. Let $D = Y^T Y/n$ and $Q_1 = [1,0,...0]^T$. We first initialize for $\theta_i^{(0)}, i = 1,...,d$:

$$\tilde{\theta}_i^{(0)} = (I - Q_i Q_i^T D)\theta_*, \quad \theta_i^{(0)} = \frac{\tilde{\theta}_i^{(0)}}{\sqrt{\tilde{\theta}_i^{T(0)} D \tilde{\theta}_i^{(0)}}}, \quad Q_{i+1} = [Q_i : \theta_i], \quad \text{where } \theta_* \text{ is a random } k \text{ - vector.}$$

Step 2: Iterate between $\theta^{(s)}$ and $\beta^{(s)}$ until convergence or until a specified maximum number of iterations (s=1,2,…)is reached:

Step A: For fixed $\theta_i^{(s-1)}, i = 1,...,d,$ we solve the following minimization problem:

$$\min_{\beta_i^{(s)} \in R^P} \sum_{i=1}^{d} ||Y\theta_i^{(s-1)} - X\beta_i^{(s)}||_2^2 + \lambda \sum_{l=1}^{p} ||\beta_l^*||_2^{1-r}, \text{ where } B^{(s)} = [\beta_1^{(s)},...\beta_d^{(s)}] = [\beta_1^{*(s)},...\beta_p^{*(s)}]. \quad (4a)$$

Step B: For fixed $\beta_i^{(s)}, i = 1,...,d$, we seek $\theta_i^{(s)}, i = 1,...,d$, which solve the following unconstrained optimization problem:

$$\min_{\theta_i^{(s)} \in R^k} ||Y\theta_i^{(s)} - X\beta_i^{(s)}||_2^2$$
$$\text{s.t.} \quad \theta_i^{(s)T} Y^T Y \theta_i^{(s)} = 1, \theta_i^{(s)T} Y^T Y \theta_j^{(s)} = 0, j = 1,...i-1.$$

Solution to the above optimization leads to a nonlinear equation:

$$\theta_i^{(s)} = \frac{(I - Q_i^{(s)} Q_i^{(s)T} D) D^{-1} Y^T X \beta_i^{(s)}}{n \theta_i^{T(s)} Y^T X \beta_i^{(s)}}. \quad (4b)$$

By Newton method, we obtain a solution $\tilde{\theta}_i^{(s)}$ to equation (4b). Set $\theta_i^{(s)} = \frac{\tilde{\theta}_i^{(s)}}{\sqrt{\tilde{\theta}_i^{T(s)} D \tilde{\theta}_i^{(s)}}}$.

## Results

To illustrate its application to selection of clinically useful genetic variants, the proposed method was first applied to several simulation datasets. Fourier series were used as the transformation function. The number of Fourier function in the simulations was 30. We used the true positive rate (TPR), defined as the proportion of the correctly identified predictors, to measure how well the method selects the predictors. We considered two scenarios: 50 SNPs and 100 SNPs SNPs from chromosome 1 in the NHLBI's Exome Sequencing Project (ESP) dataset with 5,406 individuals and 1,779,016 SNPs. The simulation models were given by

$$y_i = \sum_{l=1}^{p} x_{ij} b_j + 0.1 \varepsilon_i, \text{ where } \varepsilon_i \text{ is distributed as a standard normal distribution } N(0,1).$$

The results were summarized in Table 1.

Table 1. True positive rates for two simulated data.

| Dataset | Sample Size | Number of SNPs | Number of Causal SNPs | TPR |
|---------|-------------|----------------|-----------------------|------|
| 1 | 5406 | 50 | 10 | 100% |
| 2 | 5406 | 100 | 10 | 100% |

To further evaluate its performance, the proposed method was also applied to the real NHLBI's Exome Sequencing Project (ESP) dataset with HDL phenotype. We discovered 863 SNPs that contribute the HDL variation. The top 10 selected SNPs were listed in Table 2. It is reported that the gene cholesteryl ester transfer protein, plasma (CETP) is associated with HDL (Braun et al. 2012), CD36 is associated with atherosclerotic cardiovascular diseases (Yuasa-Kawase et al. 2012),

Table 2. Top 10 selected SNPs

| CHR | SNP | Gene | P-value | CHR | SNP | Gene | P-value |
|-----|-----|------|---------|-----|-----|------|---------|
| chr16 | rs34065661 | CETP | 9.624E-17 | chr19 | rs1052983 | LILRA6 | 1.57313E-07 |
| chr7 | rs3211938 | CD36 | 4.629E-13 | chr7 | rs10085732 | | 1.66681E-07 |
| chr6 | rs17622 | DDO | 4.333E-08 | chr19 | rs1868953 | | 3.45063E-07 |
| chr8 | rs111855567 | | 5.861E-08 | chr19 | rs117156027 | | 3.45063E-07 |
| chr1 | rs12088246 | PTGFR | 1.512E-07 | chr1 | rs79907831 | SPOCD1 | 1.15888E-06 |